\documentclass[preprint,12pt]{elsarticle}
\usepackage{amssymb}
\usepackage{amsmath}
\usepackage{graphicx,subfig}
\usepackage{booktabs}
\usepackage{subcaption}
\let\today\relax
\makeatletter
\def\ps@pprintTitle{%
	\let\@oddhead\@empty
	\let\@evenhead\@empty
	\def\@oddfoot{\footnotesize\itshape
		{Submitted preprint} \hfill\today}%
	\let\@evenfoot\@oddfoot
}
\makeatother

\begin{document}

\begin{frontmatter}
\author[aff1]{Stefano Della Fiore}
\author[aff1]{Alessandro Gnutti}
\author[aff1]{Marco Dalai}
\author[aff1]{Pierangelo Migliorati}
\author[aff1]{Riccardo Leonardi}

\affiliation[aff1]{organization={Department of Information Engineering, University of Brescia},
             country={Italy}}
 \title{End-to-End Semantic Preservation in Text-Aware Image Compression Systems\tnoteref{label1}}
\tnotetext[label1]{This work was supported in part by the European Union under the Italian National Recovery and Resilience Plan (NRRP) of NextGenerationEU, Partnership on ``Telecommunications of the Future," Program ``RESTART” under Grant PE00000001, ``Netwin” Project (CUP E83C22004640001) and ``FRAME" Project (CUP C89J24000250004). An earlier version of this paper was presented in part at EUSIPCO 2025.}

\begin{abstract}
Traditional image compression methods aim to reconstruct images for human perception, prioritizing visual fidelity over task relevance. In contrast, Coding for Machines focuses on preserving information essential for automated understanding. Building on this principle, we present an end-to-end compression framework that retains text-specific features for Optical Character Recognition (OCR). The encoder operates at roughly half the computational cost of the OCR module, making it suitable for resource-limited devices. When on-device OCR is infeasible, images can be efficiently compressed and later decoded to recover textual content. Experiments show significant improvements in text extraction accuracy at low bitrates, even outperforming OCR on uncompressed images.

We further extend this study to general-purpose encoders, exploring their capacity to preserve hidden semantics under extreme compression. Instead of optimizing for visual fidelity, we examine whether compact, visually degraded representations can retain recoverable meaning through learned enhancement and recognition modules. Results demonstrate that semantic information can persist despite severe compression, bridging text-oriented compression and general-purpose semantic preservation in machine-centered image coding.
\end{abstract}


\end{frontmatter}

\section{Introduction}
For decades, image compression has been a crucial research area, driven by the increasing demand for efficient data storage and transmission across diverse digital applications. The advent of deep learning has transformed this field, driving significant interest in end-to-end learned compression frameworks, with variational autoencoder (VAE)-based methods playing a key role~\cite{balle2016end,balle2018variational}. Unlike traditional techniques, these approaches jointly optimize the encoding and decoding stages within a unified framework. Recent works~\cite{zou2022devil,zhu2022transformerbased, chen2022two,he2022elic,wang2022neural,Liu_2023_CVPR} have demonstrated that learned compression techniques can outperform VVC~\cite{vvc}, the state-of-the-art standard for image and video coding, in terms of key quality metrics such as Peak Signal-to-Noise Ratio (PSNR) and Multi-Scale Structural Similarity (MS-SSIM). These findings underscore the transformative potential of learned approaches in shaping the future of image coding~\cite{jpegai}.

Although most learned image compression methods are tailored for human perception, the growing prominence of machine vision applications has driven a shift toward task-specific optimization. Recently, image coding for machine perception has gained significant attention, fueled by the increasing need to transmit visual data efficiently for high-level recognition tasks across devices. Referred to as Image Coding for Machines (ICM), this strategy aims to refine the compression process to generate images better suited for specific downstream tasks. Some examples of tasks in this domain include denoising~\cite{pcs24}, computer vision tasks~\cite{Chen_2023_ICCV}, face detection in the compressed domain~\cite{mmsp24}, and, more recently, Multimodal Large Language Models (MLLMs)~\cite{kao2025}.

Among these recognition applications, Optical Character Recognition (OCR) is a technology that enables the automatic extraction of text from images, scanned documents, and other visual data sources. OCR systems can convert printed or handwritten text into machine-readable formats. This capability is widely used across various applications, including document digitization, automated data entry, and assistive technologies for the visually impaired. Modern OCR systems often employ deep neural networks to enhance accuracy, enabling them to handle diverse fonts, handwriting styles, and challenging conditions such as poor lighting or distortions.

\begin{figure}[t]
\centering
	\includegraphics[width=0.8\linewidth]{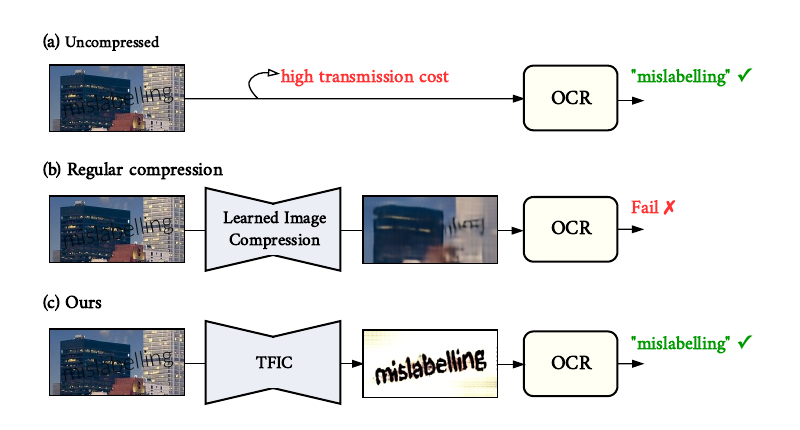}
\caption{Comparison of high-level frameworks: no compression, conventional compression, and our proposed TFIC.}
\label{fig:high-level}
\end{figure}

However, a major challenge arises when applying image compression to OCR-relevant images. Compression techniques, especially at low bitrates, can introduce significant distortions, such as blurring, blocking artifacts, and loss of fine details, which can severely impact text readability. Since OCR performance heavily depends on the clarity and integrity of textual structures, excessive compression may lead to character misinterpretation or complete recognition failure.

In this paper, we introduce an image compression system designed specifically for OCR-based vision tasks. Our learnt Text-Focused Image Compression (TFIC) model (see Figure \ref{fig:high-level}) is built on a standard architecture commonly used for neural image compression. However, we incorporate an OCR model at the system's output and optimize the compression process by minimizing the loss specific to OCR performance. This approach allows us to achieve a balance between efficient compression and high-quality text recognition, even at extremely lower compression rates where traditional methods strongly degrade the text to the point of being unreadable.

Then, we further extend this study to general-purpose encoders, exploring their capacity to preserve hidden semantics under extreme compression. Instead of optimizing for visual fidelity, we examine whether compact, visually degraded representations can retain recoverable meaning through learned enhancement and recognition modules.

The remainder of the paper is organized as follows. Section~\ref{sec:background} provides an overview of neural image compression and OCR systems. Section~\ref{sec:proposed-method} outlines the architecture of our method and its end-to-end training approach and presents experimental results and comparisons with baseline methods. 
Finally, in Section~\ref{sec: hiddensem}, we extend our analysis to the use of general-purpose image encoders, investigating their behavior outside conventional operating conditions. 
Specifically, we explore scenarios in which the goal is not merely the preservation of visual fidelity, but rather the retention and recovery of \emph{semantic content} under extreme compression settings. 
This part of our study examines whether it is possible to exploit highly compressed and visually degraded image representations, often appearing corrupted or unintelligible to the human eye, while still be able to recover meaningful information through learned enhancement and recognition modules. 
The motivation for this investigation arises from the limitations of traditional compression algorithms, such as JPEG, which are optimized for perceptual quality at standard bitrates but apparently fail to preserve semantics when operating at extremely low bitrates. 
By analyzing and modeling this behavior, we are going to prove that, even under severe quantization, traces of semantic structure may persist within the compressed domain and can be effectively recovered using deep learning–based post-decoding systems.

\section{Background}
\label{sec:background}
\subsection{Neural image compression}
\label{subsec:neural-image-compression}
An end-to-end learned image compression system generally consists of two key components: the main autoencoder and the hyperprior autoencoder. The main autoencoder is composed of an analysis transform $g_a$ and a synthesis transform $g_s$. The analysis transform $g_a$ encodes an RGB image $x \in \mathbb{R}^{H \times W \times 3}$, with height $H$ and width $W$, into a latent representation $y \in \mathbb{R}^{\frac{H}{16} \times \frac{W}{16} \times C_y}$ using an encoding distribution $q_{g_a}(y|x)$. This latent $y$ is then uniformly quantized as $\hat{y}$ and entropy encoded into a bitstream via a learned prior distribution $p(\hat{y})$. On the decoder side, $\hat{y}$ is entropy decoded and reconstructed as $\hat{x} \in \mathbb{R}^{H \times W \times 3}$ using a decoding distribution $q_{g_s}(\hat{x}|\hat{y})$, implemented by the synthesis transform $g_s$. Throughout this process, the prior distribution $p(\hat{y})$ plays a critical role in determining the number of bits needed to represent the quantized latent $\hat{y}$. To mitigate this, a hyperprior autoencoder~\cite{balle2018variational} is introduced to model the prior distribution in a content-adaptive manner. The hyperprior autoencoder consists of a hyperprior analysis transform $h_a$ and a hyperprior synthesis transform $h_s$. The transform $h_a$ converts the image latent $y$ into side information $z \in \mathbb{R}^{\frac{H}{64} \times \frac{W}{64} \times C_z}$, which typically represents a small portion of the compressed bitstream. This quantized version of $z$ is then decoded from the bitstream through $h_s$, resulting in the learned prior distribution $p(\hat{y})$.

\subsection{OCR systems}
\label{subsec:ocr-systems}
An OCR system is designed to extract text from images captured in unconstrained environments. The system generally consists of four integrated modules. First, a \emph{detection} module identifies and localizes regions containing text within an image, often using object detection frameworks to output bounding boxes around potential text areas~\cite{baek2019character, he2017single, zhou2017east}. Next, the \emph{transformation} stage normalizes these detected regions to correct for distortions such as skew, rotation, or perspective changes. For instance, a Spatial Transformer Network (STN) with a Thin-Plate Spline (TPS)~\cite{STN} transformation may be employed to obtain a rectified image, denoted as $\tilde{x}$.

Subsequently, a \emph{feature extraction} component, typically implemented with a convolutional neural network (CNN) like ResNet~\cite{ResNet}, converts the (transformed) image $x$ or $\tilde{x}$ into a rich feature map $V = \{v_i\}_{i=1}^{I}$, where each $v_i$ represents features corresponding to a specific region of the image and $I$ is the dimension of the feature map. Given the sequential nature of text, these features are then reorganized into a sequence and processed by a \emph{sequence modeling} module—commonly a bidirectional LSTM (BiLSTM) (see~\cite{CRNN,RARE,FAN}) or Transformer—that captures contextual dependencies between characters, yielding a contextualized representation. Finally, the \emph{prediction} stage decodes this representation into the final text output using methods such as Connectionist Temporal Classification (CTC)~\cite{CTC} or attention-based decoding~\cite{RARE, FAN}. The entire pipeline is trained end-to-end by minimizing a loss function that measures the discrepancy between the predicted sequence and the ground truth, ensuring robust performance in complex and real-world scenarios.

\section{Proposed Text-Focused Image Compression}
\label{sec:proposed-method}
\subsection{Architecture Overview}

\begin{figure*}[t]
\centering
	\includegraphics[width=1.0\textwidth]{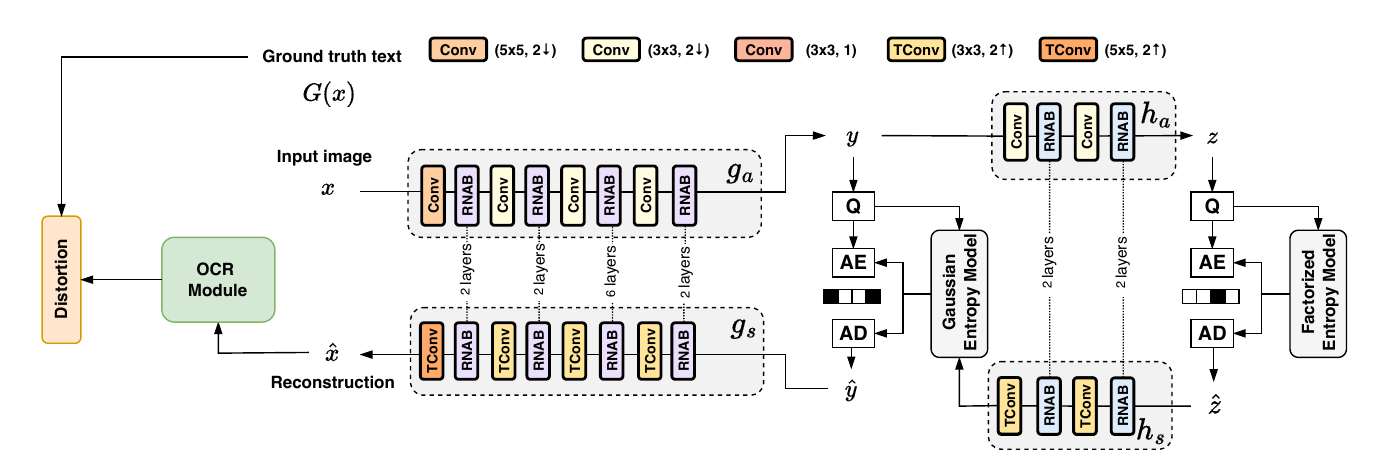}
\caption{High-level architectural framework of TFIC. The main image codec is composed by an encoder $g_a$ that compresses the input image into a latent representation, which is quantized and transmitted. The decoder $g_s$ reconstructs the image that is subsequently processed by the OCR module for text extraction.}
\label{fig:overview}
\end{figure*}

Figuree~\ref{fig:overview} illustrates our proposed framework. The image codec architecture follows the key modules outlined in~\ref{subsec:neural-image-compression}. As a reference, we adopt the image compression model from~\cite{lu2022high}, which utilizes Transformer-based backbones to implement both the main and hyperprior autoencoders.

An OCR system with frozen parameters is integrated downstream of the decoder alongside the autoencoder. Its architecture follows the key modules outlined in~\ref{subsec:ocr-systems}, with the Scene Text Recognition model from~\cite{baek2019wrong} serving as a reference. During training, the OCR extracts text from $\hat{x}$, denoted as $T(\hat{x})$, which is compared to the ground truth text $G(x)$. The resulting loss signal is then backpropagated through the entire network. This process helps guide both the encoder and decoder to preserve text-relevant information.

\subsection{Loss Function and End-to-End Training}
The overall training objective $\mathcal{L}_{\text{total}}$ of our method is a combination of three loss components:
\begin{equation*}
    \mathcal{L}_{\text{total}} = \lambda \cdot \mathcal{L}_{\text{dist}}(x,\hat{x}) + \mathcal{L}_{\text{rate}}(\hat{y},\hat{z}) + \gamma \cdot \mathcal{L}_{\text{OCR}}(G(x), T(\hat{x})),
\end{equation*}
where:
\begin{itemize}
    \item $\mathcal{L}_{\text{dist}}(x,\hat{x})$ is the distortion loss, measured by the mean squared error (MSE) between the original image $x$ and its reconstruction $\hat{x}$.
    \item $\mathcal{L}_{\text{rate}}(\hat{y},\hat{z})=-\log p(\hat{z})-\log p(\hat{y}|\hat{z})$ represents the rate loss, which estimates the bitstream length needed to encode the quantized latent representation $\hat{y}$ and side information $\hat{z}$.
    \item $\mathcal{L}_{\text{OCR}}(G(x), T(\hat{x}))$ is the OCR loss that quantifies the difference between the ground truth text $G(x)$ and the text recognized from the reconstructed image $T(\hat{x})$. This is measured using the cross-entropy loss between the text-indexes of the ground truth text and the reconstructed one.
\end{itemize}

The parameters $\lambda$ and $\gamma$ balance the trade-offs between pixel-level fidelity, compression rate, and text recognition accuracy. Our training procedure consists of two stages. In the first stage, we set $\gamma=0$, allowing the model to be pre-trained solely with the MSE for a given target rate, which is defined by the value of $\lambda$. In the second stage, we set $\lambda=0$, enabling the model to minimize only the OCR loss. During this stage, $\gamma$ is set to 0.1 for all rate points.

Training is conducted end-to-end, enabling the encoder, decoder to co-adapt such that the final compressed representation retains essential textual information while achieving high compression efficiency. We recall that only the image codec is re-trained, while the OCR system remains unchanged.

\begin{figure}[t]
\centering
\includegraphics[width=0.55\linewidth]{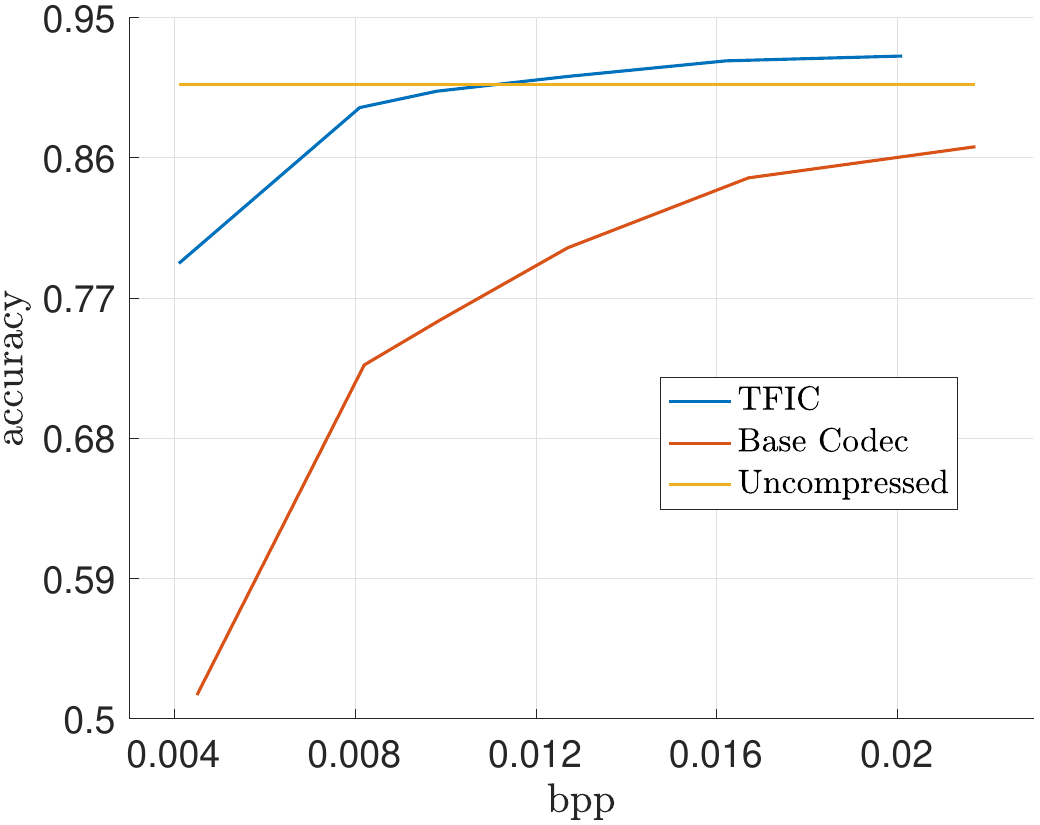}
\caption{Rate vs. OCR performance curves for images that are uncompressed, decoded with the pre-trained MSE-based codec, and decoded using the proposed TFIC.}
\label{fig:results}
\end{figure}

\begin{figure*}[t]
\centering
\subfloat[][Original Image \vspace{0.1cm} \\ \scriptsize Bpp: 0.36 \\ Accuracy: 48\%]{\setlength{\fboxsep}{0pt} \fbox{\includegraphics[width=0.23\textwidth]{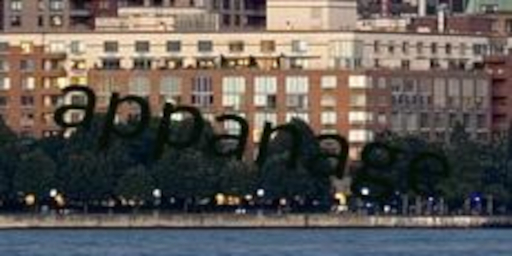}}}\
\subfloat[][Base codec  \vspace{0.1cm} \\ \scriptsize Bpp: 0.016 \\ Accuracy: 19\%]{\setlength{\fboxsep}{0pt} \fbox{\includegraphics[width=0.23\textwidth]{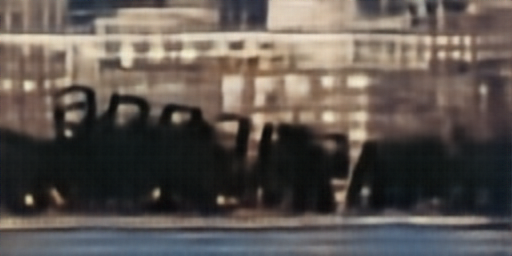}}}\
\subfloat[][Base codec \vspace{0.1cm} \\ \scriptsize Bpp: 0.0082 \\ Accuracy: 3\%]{\setlength{\fboxsep}{0pt} \fbox{\includegraphics[width=0.23\textwidth]{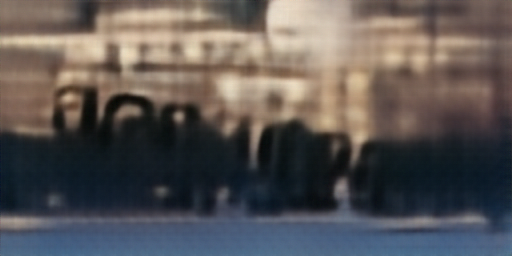}}}\
\subfloat[][TFIC  \vspace{0.1cm} \\ \scriptsize Bpp: 0.0080 \\ Accuracy: 100\%]{\setlength{\fboxsep}{0pt} \fbox{\includegraphics[width=0.23\textwidth]{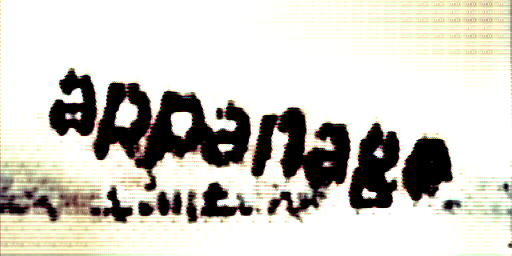}}}\\
\subfloat[][Original Image \vspace{0.1cm} \\ \scriptsize Bpp: 0.45 \\ Accuracy: 70\%]{\setlength{\fboxsep}{0pt} \fbox{\includegraphics[width=0.23\textwidth]{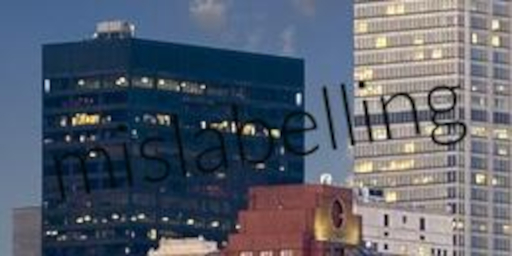}}}\
\subfloat[][Base codec \vspace{0.1cm} \\ \scriptsize Bpp: 0.016 \\ Accuracy: 42\%]{\setlength{\fboxsep}{0pt} \fbox{\includegraphics[width=0.23\textwidth]{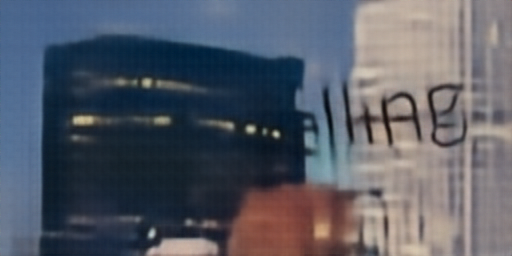}}}\
\subfloat[][Base codec \vspace{0.1cm} \\ \scriptsize Bpp: 0.0082 \\ Accuracy: 2\%]{\setlength{\fboxsep}{0pt} \fbox{\includegraphics[width=0.23\textwidth]{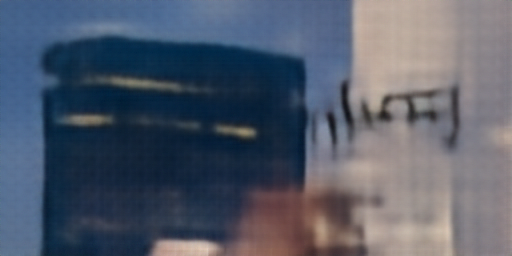}}}\
\subfloat[][TFIC \vspace{0.1cm} \\ \scriptsize Bpp: 0.0080 \\ Accuracy: 100\%]{\setlength{\fboxsep}{0pt} \fbox{\includegraphics[width=0.23\textwidth]{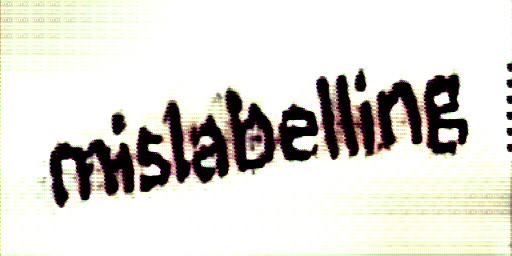}}}\\
\subfloat[][Original Image \vspace{0.1cm} \\ \scriptsize Bpp: 0.37 \\ Accuracy: 100\%]{\setlength{\fboxsep}{0pt} \fbox{\includegraphics[width=0.23\textwidth]{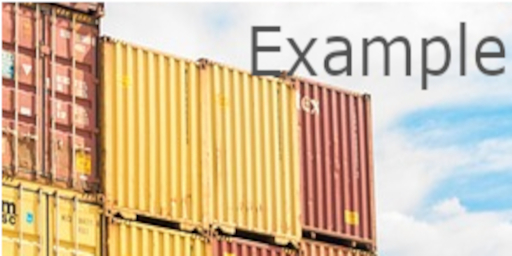}}}\
\subfloat[][Base codec \vspace{0.1cm} \\ \scriptsize Bpp: 0.016 \\ Accuracy: 100\%]{\setlength{\fboxsep}{0pt} \fbox{\includegraphics[width=0.23\textwidth]{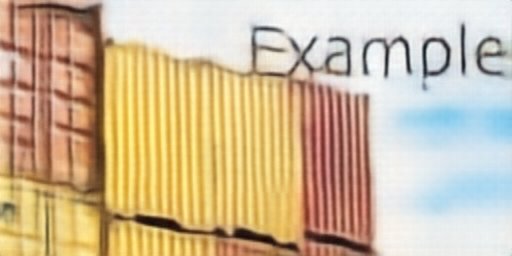}}}\
\subfloat[][Base codec \vspace{0.1cm} \\ \scriptsize Bpp: 0.0082 \\ Accuracy: 80\%]{\setlength{\fboxsep}{0pt} \fbox{\includegraphics[width=0.23\textwidth]{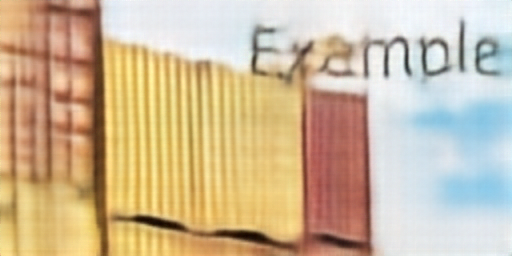}}}\
\subfloat[][TFIC \vspace{0.1cm} \\ \scriptsize Bpp: 0.0080 \\ Accuracy: 100\%]{\setlength{\fboxsep}{0pt} \fbox{\includegraphics[width=0.23\textwidth]{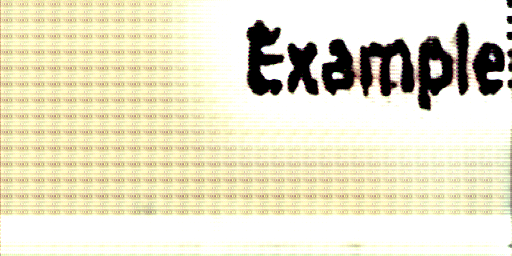}}}
\caption{Visual comparison of reconstructed images. Although the base codec preserves more global details, our method retains critical textual regions that yield superior OCR performance.}
\label{fig:visual-results}
\end{figure*}

\begin{figure}[t]
\centering
\includegraphics[width=0.55\linewidth]{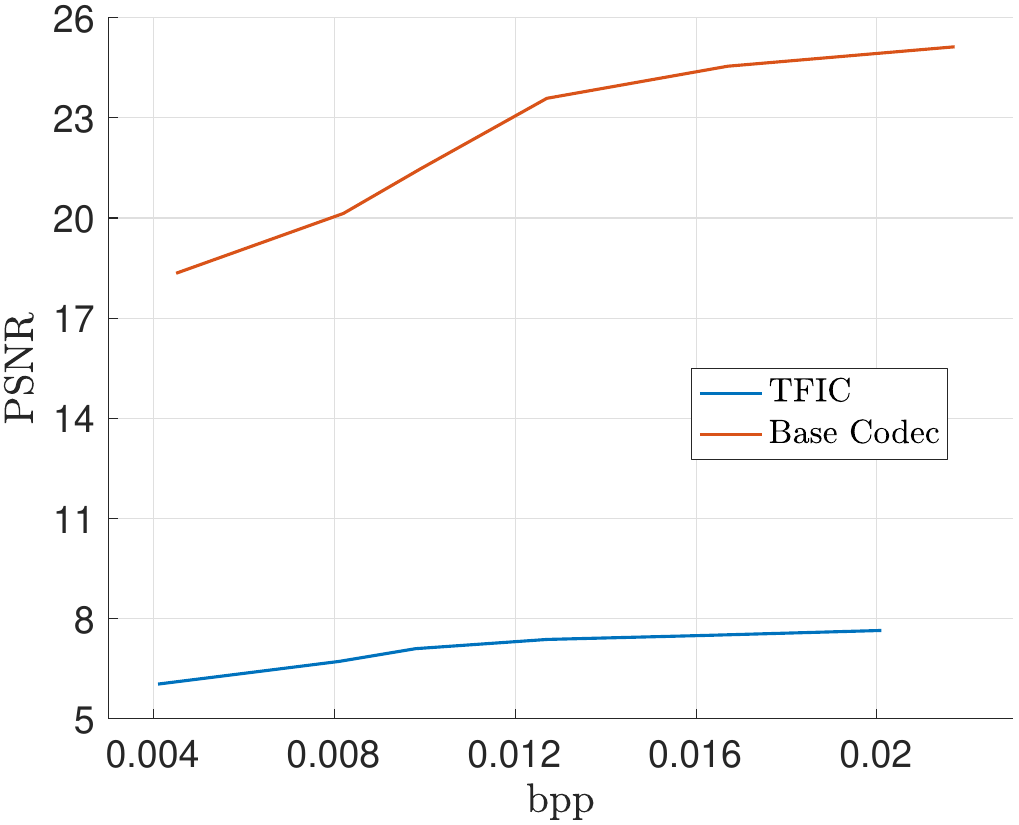}
\caption{Rate vs. PSNR performance curves for the base codec and the proposed TFIC.}
\label{fig:resultsPSNR}
\end{figure}

\subsection{Dataset and Experimental Setup}
We use a synthetic data generator for text recognition~\cite{textgenerator} to create training and test sets of approximately 20k and 600 images, respectively. Each image is resized to a fixed resolution of $256 \times 512$ pixels. The dataset covers a diverse range of fonts, layouts, and background complexities, ensuring a comprehensive evaluation of text extraction under various conditions.

For evaluation, we compare the proposed method with the base codec optimized exclusively for MSE on the same training set, thus without including the OCR loss component. Both models are evaluated based on bitrate (bits-per-pixel, bpp) and OCR performance, which is measured by the Levenshtein edit-distance $\text{lev}(G(x), T(\hat{x}))$ (see~\cite{levenshtein1966binary}) that represents the minimum number of single-character edits (insertions, deletions or substitutions) required to change the reconstructed text $T(\hat{x})$ into the ground truth one $G(x)$. The accuracy is then defined as
$$
    1 - \frac{\text{lev}(G(x), T(\hat{x}))}{\max\left\{ |G(x)|, |T(\hat{x})|  \right\}}\,,
$$
where $|G(x)|$ and $|T(\hat{x})|$ are the number of alphanumeric characters contained in the ground truth text and in the reconstructed one, respectively.

\subsection{Results}
Figure~\ref{fig:results} shows the rate vs. OCR performance curves for both the base codec and the proposed TFIC model. The base codec exhibits a significant drop in OCR accuracy, with lower accuracy values at comparable bitrates. In contrast, our method consistently achieves higher accuracy, effectively preserving text information even at lower bitrates. Notably, TFIC even surpasses the OCR performance on uncompressed images, suggesting that it acts as a beneficial pre-processing step.

This observation is supported by Figure~\ref{fig:visual-results}, which provides a visual comparison of reconstructed images from the base codec and TFIC. While the base codec achieves higher overall PSNR (see Figure~\ref{fig:resultsPSNR}), text regions often appear blurred or distorted, resulting in poor OCR accuracy. In contrast, our method preserves sharp and legible text, even if non-essential areas of the image undergo destructive compression.

A comparison of the results shown in the second, third and fourth columns of Figure ~\ref{fig:visual-results}, shows that our method can also be useful when used \emph{in addition} to a standard compression scheme, whenever transmission of a low rate visual content of the background is also required. Indeed, the total rate for sending \emph{both} images in the third and fourth columns (as two separate layers) equals the rate for the image in the second column. The computation time would in this case be doubled, but would still remain lower than the time needed for the OCR on the original image alone, while also doing compression.

\begin{table}[t]
\normalsize
\centering
\begin{tabular}{@{}cccc@{}}
\toprule
 & Encoding & OCR module \\ \midrule
Computational times & $12.9 \pm 1.8$ & $24.1 \pm 3.3$ \\ \bottomrule
\end{tabular}
\caption{Average and standard deviation of encoding and OCR times per image, measured in milliseconds.}
\label{tab:metrics}
\end{table}

\subsection{Runtime Analysis}
One motivation behind our method is a difference in processing time between the encoding and OCR stages. Empirical measurements indicate that the encoding process of our method requires only about one half of the time needed to perform OCR on the same image. This efficiency is crucial for resource-limited devices, enabling rapid on-device compression with deferred, more intensive OCR processing on external hardware. Table~\ref{tab:metrics} shows the average and standard deviation of encoding and OCR times per image computed over the entire test set, revealing that the encoding stage takes roughly half the time required by the OCR stage.

\section{Hidden Semantic Preserved by General-Purpose Encoders}\label{sec: hiddensem}

In the second phase of our study, we expanded our investigation to the use of general-purpose encoders, particularly focusing on how these encoders behave outside their standard operating ranges when the objective is not mere visual fidelity but rather the preservation of semantic content.
The underlying question is whether it is possible to exploit extremely compressed representations, although corrupted and unintelligible in a conventional sense, recovering meaningful semantic information through learned enhancement and recognition modules.

\subsection{Motivation}

While traditional compression systems, such as JPEG, are designed to minimize perceptual distortion at typical operating points, they are not tailored to preserve semantic content at extremely low bitrates. In scenarios such as large-scale data transmission, where bandwidth is critically limited, operating at these extreme compression levels may be unavoidable. The main challenge is that such aggressive quantization typically removes most of the low-level features necessary for human or machine interpretation. As a result, the reconstructed images appear as noisy, blurred, or completely corrupted patterns.

Our work is motivated by the observation that semantic information, especially in the case of text, may still be implicitly encoded within these degraded signals, even if the content is visually incomprehensible. The goal is then to design post-decoder enhancement systems capable of leveraging prior knowledge, through deep learning models, to recover this semantic information.

\begin{figure}[t]
\centering
	\includegraphics[width=0.95\linewidth]{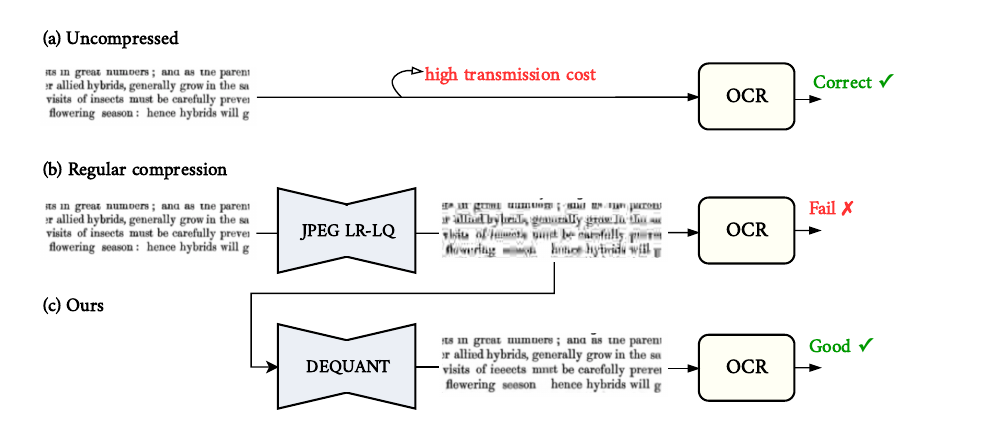}
\caption{Comparison of high-level frameworks: no compression, regular compression, and our proposed method.}
\label{fig:high-level-quant}
\end{figure}

\subsection{Case Study: Textual Images}

As a case use, we focused on textual images subjected to extremely low-rate JPEG compression. The dataset consisted of low-resolution digital text images and these images were compressed using JPEG with extreme quantization levels, leading to a representation that to the human eye resembles random noise or blotches of black and white pixels. Unsurprisingly, applying a conventional OCR system to these compressed images yields recognition rates close to zero, confirming that no usable information can be extracted through standard pipelines.

Despite this, we hypothesized that the statistical correlations present in such degraded images could still retain implicit traces of the underlying semantic content. To exploit this, we trained convolutional neural networks specifically designed to map compressed JPEG inputs back to a legible textual domain.

In Figure~\ref{fig:high-level-quant}, a high-level diagram of our method is presented. The block JPEG LR-LQ represents a module that performs JPEG compression on low-resolution images using extremely high quantization, leading to significant compression and loss of detail. The block DEQUANT denotes a convolutional module designed to perform dequantization, effectively learning to restore image details that have been degraded by the high quantization in the JPEG compression stage.

The first step involved designing a convolutional neural network to post-process the decoded JPEG images. Even with a simple mean squared error (MMSE) objective function, the network was able to restore structural features of the characters, producing outputs sufficiently clear for OCR modules to operate. This result highlights that even when low-level image fidelity is lost, mid-level structures associated with characters remain recoverable.

The architecture adopted in our preliminary experiments consisted of a lightweight convolutional neural network with residual connections, allowing the model to focus on refining corrupted strokes and suppressing JPEG artifacts. Interestingly, the restored images did not always visually resemble the originals in terms of fine details; however, they were interpretable at a semantic level, sufficient for both OCR recognition and human readability.

\subsection{Architecture Overview}
\label{sec:architecture}

\begin{figure}[t]
    \centering
    \includegraphics[width=0.7\linewidth]{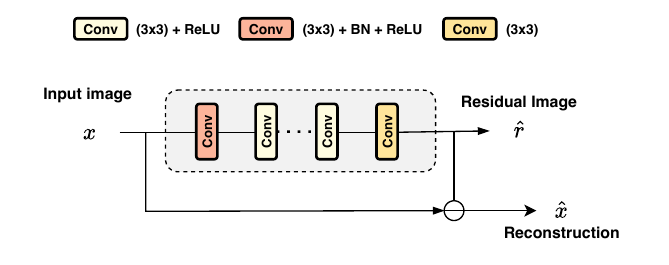}
    \caption{High-level schematic representation of the proposed DnCNN-based architecture used for image dequantization. The model predicts the residual component, which is subtracted from the input to obtain the restored image.}
    \label{fig:architecture}
\end{figure}

The proposed network is based on a modified DnCNN architecture designed to perform image restoration and dequantization on JPEG-compressed low-resolution patches. 
As illustrated in Figure~\ref{fig:architecture}, the model follows a fully convolutional design composed of a series of feature extraction, normalization, and residual learning blocks.

The network receives as input an image patch of size $64 \times 256$ and processes it through an initial convolutional layer with $64$ filters of size $3\times3$, followed by a ReLU activation. 
Subsequently, a sequence of $(D-2)$ convolutional blocks, each composed of a $3\times3$ convolution, batch normalization (BN), and ReLU activation, progressively refines the feature representations. 
In our implementation, we set the network depth to $D = 17$, resulting in $15$ intermediate convolutional layers with batch normalization. 
The final convolutional layer produces a residual output $\hat{r}$ of the same spatial dimensions as the input.

The network is trained to estimate the residual component (i.e., the quantization noise) rather than the clean image directly. 
The restored image $\hat{x}$ is obtained as
$$
\hat{x} = x - \hat{r},
$$
where $x$ denotes the input patch and $\hat{r}$ is the predicted noise component.

\subsection{Loss Function}

The training objective is defined as a combination of a pixel-domain reconstruction loss and a frequency-domain consistency loss. The overall loss function is expressed as
$$
\mathcal{L}_{\text{total}} = \lambda \cdot \mathcal{L}_{\text{dist}}(x, \hat{x}) + (1-\lambda) \cdot \mathcal{L}_{\text{DCT}}(x, \hat{x}),
$$
where $\lambda \in [0,1]$ is a weighting coefficient that balance the contribution of each term.

The first term, $\mathcal{L}_{\text{dist}}(x, \hat{x})$, corresponds to the classical mean squared error loss between the reconstructed image $\hat{x}$ and the original image $x$ which enforces pixel-wise fidelity in the spatial domain.

The second term, $\mathcal{L}_{\text{DCT}}(x, \hat{x})$, is a blockwise Discrete Cosine Transform (DCT) loss designed to enforce frequency-domain consistency between the reconstructed image $\hat{x}$ and the input quantized image $x$. Let $\mathcal{B}$ denote the set of non-overlapping $8\times8$ blocks extracted from the images $x$ and $\hat{x}$. For each block $b \in \mathcal{B}$, let $x_b, \hat{x}_b \in \mathbb{R}^{8\times8}$ denote the corresponding image patches. Then, the two-dimensional DCT is applied to each block, following the standard JPEG formulation. The resulting DCT coefficients are compared through a thresholded absolute difference:
$$
\mathcal{L}_{\text{DCT}}(x, \hat{x}) = \mathbb{E}_{b \sim \mathcal{B}}\left[\max\left(|DCT(x_b) - DCT(\hat{x}_b)| - \tau,\, 0\right)\right],
$$
where $\tau=0.5$ is a fixed margin that suppresses small, perceptually irrelevant differences. This formulation helps the reconstructed image to remain coherent with the quantized input in the frequency domain, ensuring that the reconstructed output does not deviate from the spectral structure imposed by the JPEG compression process.

The combination of $\mathcal{L}_{\text{dist}}$ and $\mathcal{L}_{\text{DCT}}$ enables the network to jointly optimize for spatial accuracy and frequency-domain consistency, leading to reconstructions that are both visually faithful and perceptually coherent.

After a preliminary analysis, we observed that the best reconstruction quality was achieved by setting the weighting coefficient to $\lambda= 0.8$.

\subsection{Extended Architecture with OCR-Guided Loss}
\label{sec:ocr-architecture}

\begin{figure}[t]
    \centering
    \includegraphics[width=0.95\linewidth]{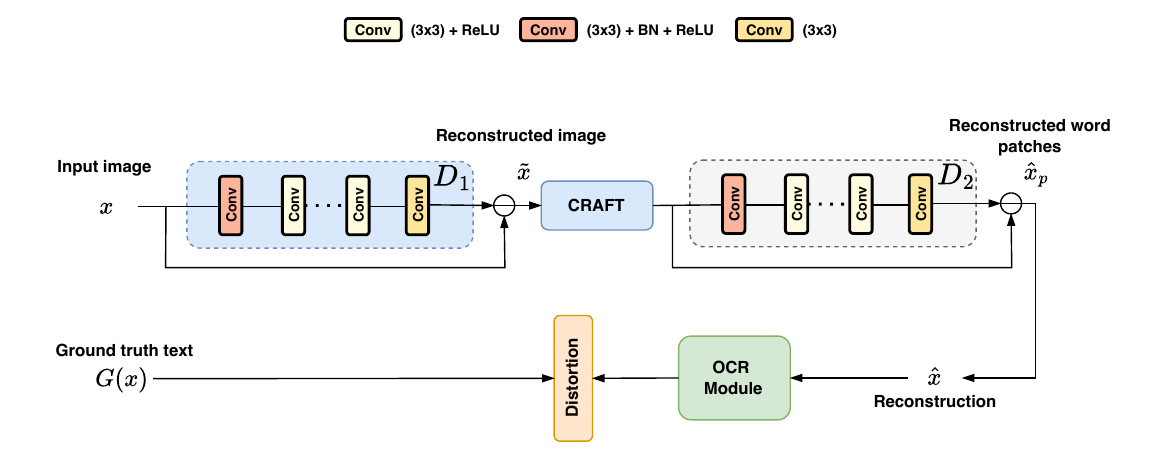}
    \caption{Schematic representation of the proposed OCR-guided dequantization architecture. 
    A frozen DnCNN produces an initial reconstruction, which is analyzed by a CRAFT detector to extract word-level patches. 
    Each patch is refined by a trainable DnCNN and evaluated by a fixed OCR module that provides a recognition-based loss for training.}
    \label{fig:ocr_architecture}
\end{figure}

To further enhance the perceptual and semantic quality of the reconstructed text regions, we extend the baseline DnCNN-based model by introducing an additional \emph{OCR-guided loss} through an integrated text understanding pipeline. 
The proposed architecture is schematically illustrated in Figure~\ref{fig:ocr_architecture}.

Given an input degraded image $x$ (affected by compression or quantization artifacts), the first stage consists of a frozen DnCNN network $D_1$, pre-trained according to the method described in Section~\ref{sec:architecture}. 
This module produces an initial reconstruction $\tilde{x} = D_1(x)$, which serves as an improved visual prior for the subsequent text analysis.

The reconstructed image $\tilde{x}$ is then processed by a freezed CRAFT (Character Region Awareness for Text Detection) network~\cite{baek2019character} which identifies and localizes individual word regions.
For each detected text region, a corresponding image patch $\tilde{x}_p$ is extracted, forming the set of word-level patches~$\mathcal{P}$.

Each patch $\tilde{x}_p$ is subsequently fed into a trainable DnCNN module $D_2$, which refines the local reconstruction:
$$
\hat{x}_p = D_2(\tilde{x}_p).
$$

The refined patch $\hat{x}_p$ is then passed to a fixed OCR recognition network placed at the end of the processing chain. 
Although the weights of the OCR module remain frozen during training, the backpropagation of the loss gradients goes through the OCR module, enabling $D_2$ to be optimized according to the semantic feedback produced by the OCR.

An OCR-based loss term $\mathcal{L}_{\text{OCR}}$ is introduced, defined as the discrepancy between the OCR predictions on the reconstructed patches and their corresponding ground-truth text labels $G_p(x)$'s.

$$
\mathcal{L}_{\mathrm{OCR}} = \mathbb{E}_{p \sim \mathcal{P}} \left[ \ell_{\text{OCR}}\!\left(\text{OCR}\!\left(D_2(\tilde{x}_p)\right), G_p(x) \right) \right]\,,
$$
where the OCR-based loss $\ell_{\text{OCR}}$ is defined as the cross-entropy between the OCR predictions and the ground-truth textual labels of each detected word.

Finally, all reconstructed patches $\hat{x}_p$ are reassembled to obtain the final restored image $\hat{x}$. 

The training phase of the proposed OCR-guided architecture was conducted starting from the pretrained DnCNN model described in Section~\ref{sec:architecture}. 
During this stage, all the parameters of the pretrained $D_1$, the CRAFT detector, and the OCR recognition module were kept frozen, while only the weights of the trainable $D_2$ were updated. Fine-tuning was performed exclusively using the OCR-based loss $\mathcal{L}_{\text{OCR}}$.
This strategy allows the network to adapt its reconstruction process to maximize the text recognition accuracy, refining the visual details specifically in regions containing textual information.

\subsection{Dataset and Experimental Results}

The dataset used for the experiments consists of text images extracted from printed books and documents. 
Each image contains black text rendered in different fonts and typographic styles on a uniform white background, reproducing realistic scanned document conditions. 
In total, $2285$ full-page images were collected.

To simulate realistic degradation, each clean image was first downscaled to a lower spatial resolution and then compressed using the JPEG codec with the highest quantization factor, corresponding to extremely aggressive compression.
This process generates low-quality, artifact-rich images characterized by severe blocking and quantization effects, which represent the degraded inputs used for network training.

To increase the dataset diversity and sample count, from each original image we randomly extracted $10$ non-overlapping patches of fixed size $64 \times 256$ pixels. 
This procedure yields a total of $22850$ image patches, each paired with its corresponding quantized version.

The dataset was subsequently divided into two disjoint subsets: 
$80\%$ of the patches were used for training, while the remaining $20\%$ were reserved for testing and validation.

In the following, we present the experimental results obtained with the proposed architectures. 
We first evaluate the baseline model, based on a single DnCNN network trained with the combined MSE and DCT losses described in Section~\ref{sec:architecture}. 
Then, we show the improvements achieved by the extended OCR-guided architecture introduced in Section~\ref{sec:ocr-architecture}.

\begin{figure*}[t]
    \centering
    \footnotesize
    \begin{tabular}{ccc}
        \includegraphics[width=0.3\linewidth]{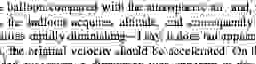} &
        \includegraphics[width=0.3\linewidth]{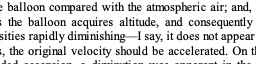} &
        \includegraphics[width=0.3\linewidth]{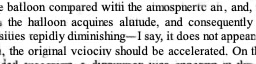} \\[1pt]
        Noisy Image &  Ground Truth &  Rec. (DnCNN) \\[1pt]
        \includegraphics[width=0.3\linewidth]{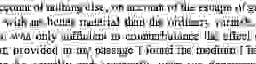} &
        \includegraphics[width=0.3\linewidth]{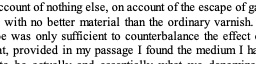} &
        \includegraphics[width=0.3\linewidth]{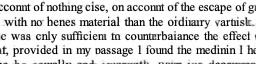} \\[1pt]
        Noisy Image & Ground Truth &  Rec. (DnCNN) \\[1pt]
        \includegraphics[width=0.3\linewidth]{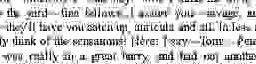} &
        \includegraphics[width=0.3\linewidth]{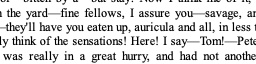} &
        \includegraphics[width=0.3\linewidth]{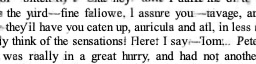} \\[1pt]
        Noisy Image & Ground Truth & Rec. (DnCNN)
    \end{tabular}
    \caption{Visual results obtained using the baseline DnCNN architecture. 
The model successfully removes quantization artifacts and reconstructs most visual details, although some text regions remain imperfectly restored. 
The OCR accuracy achieved by this model was approximately $75\%$.}
    \label{fig:dncnn_results}
\end{figure*}

The first set of experiments aims to assess the reconstruction quality achieved by the baseline DnCNN model operating directly on JPEG-compressed and quantized images.
Figure~\ref{fig:dncnn_results} shows some qualitative results.
Each example shows, from left to right: the degraded input image, the corresponding clean ground-truth, and the restored output produced by the baseline DnCNN.

As can be observed, the network effectively removes most of the compression artifacts and restores fine details such as edges and text contours. 
Although the reconstructions are visually accurate and preserve global image structure, some residual degradation can still be noticed where minor distortions or character ambiguities may remain.

\begin{figure*}[t]
    \centering
    \footnotesize
    \begin{tabular}{cc}
         \includegraphics[width=0.45\linewidth]{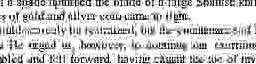} &
        \includegraphics[width=0.45\linewidth]{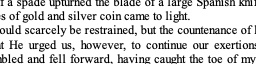}  \\[1pt]
        Noisy Image & Ground Truth \\[1pt]
        \includegraphics[width=0.45\linewidth]{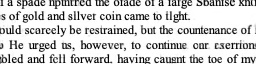} &
        \includegraphics[width=0.45\linewidth]{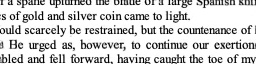} \\[1pt]
        Rec. (DnCNN) & Rec. (OCR) \\[4pt]
        \multicolumn{2}{c}{\rule{0.9\linewidth}{0.4pt}} \\[4pt]
        \includegraphics[width=0.45\linewidth]{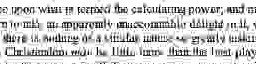} &
        \includegraphics[width=0.45\linewidth]{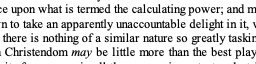}  \\[1pt]
        Noisy Image & Ground Truth \\[1pt]
        \includegraphics[width=0.45\linewidth]{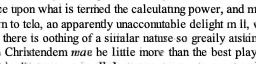} &
        \includegraphics[width=0.45\linewidth]{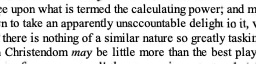} \\[1pt]
        Rec. (DnCNN) & Rec. (OCR) \\[1pt]
    \end{tabular}
    \caption{Comparison between the baseline DnCNN and the proposed OCR-guided architecture. 
The baseline DnCNN achieved an OCR accuracy of approximately $75\%$, while the proposed OCR-guided model reached about $90\%$.}
    \label{fig:ocr_results}
\end{figure*}

To address the remaining inconsistencies observed in textual areas, we further evaluated the proposed OCR-guided architecture described in Section~\ref{sec:ocr-architecture}. 
This extended model integrates semantic feedback from a frozen OCR module during training, allowing it to refine the reconstruction.

As shown in Figure~\ref{fig:ocr_results}, several qualitative differences can be observed between the reconstructions obtained using the baseline DnCNN and those produced by the OCR-guided network. 
In particular, some words reconstructed by the baseline DnCNN exhibit incorrect or malformed characters, resulting in textual sequences that do not correspond to valid words or that differ semantically from the ground truth. In contrast, the reconstructions obtained with the proposed OCR-guided approach demonstrate an improvement in textual consistency.

In addition to the qualitative results, we evaluated the OCR accuracy on a test set composed of $4570$ patches, each of size $64 \times 256$.
The OCR accuracy on the original quantized and heavily compressed images was extremely low, with an average recognition rate below $5\%$, confirming the severe degradation introduced by the high quantization levels. 
When applying the baseline DnCNN model, the accuracy increased significantly, reaching approximately a value of $75\%$, indicating that the network effectively restored most of the visual information necessary for OCR recognition. 
Finally, with the proposed OCR-guided architecture, the accuracy further improved to approximately $90\%$, demonstrating that the inclusion of the OCR-based loss allows the model to refine textual details in a semantically consistent manner, leading to substantial gains in both readability and recognition reliability.

\bibliographystyle{elsarticle-num} 
\bibliography{refs}
\end{document}